\documentstyle[emulateapj]{article}
\slugcomment{The Astrophysical Journal, 566:000-000, 2001 February 10}

\begin{document}
\def\be{\begin{equation}}
\def\ee{\end{equation}}
\def\te{t_{\rm E}}
\def\au{{\rm AU}}
\def\thetae{\theta_{\rm E}}
\def\murel{\mu_{\rm rel}}
\def\dos{D_{\rm s}}
\def\dls{D_{\rm ls}}
\def\dol{D_{l}}
\def\drel{D_{\rm rel}}
\def\days{\rm days}
\def\msun{M_\odot}
\def\muas{\mu{\rm as}}
\def\mjup{M_{\rm J}}
\def\kms{{\rm km~s^{-1}}}
\def\kpc{{\rm kpc}}
\def\pion{P_{\rm ion}}
\def\pprim{P_{\rm det}}
\def\re{R_{\rm E}}
\def\sigmap{\sigma_{\rm P}}
\def\citep#1{(\cite{#1})}
\def\eq#1{equation~(\ref{#1})} 
\def\Eq#1{Eq.~\ref{#1}}
\def\murel{\mu_{\rm rel}}
\def\dt{\Delta{t}}
\def\at{A_{\rm T}}
\def\ut{u_{\rm T}}
\def\dtil{\tilde d}
\def\ap{a_P}
\def\tenc{t_{\rm enc}}
\def\tevap{t_{\rm evap}}
\def\tex{t_{\rm ex}}
\def\ttid{t_{\rm tid}}
\def\fevap{f_{\rm evap}}
\def\fion{f_{\rm ion}}
\def\tcross{t_{\rm cross}}
\def\trelax{t_{\rm relax}}
\def\teq{t_{\rm eq}}
\def\deg{^\circ\hskip-2pt}
\def\ave#1{\left<#1\right>}

\lefthead{GAUDI} \righthead{INTERPRETING THE M22 EVENTS}

\title{Interpreting the M22 Spike Events}

\author{B. Scott Gaudi\altaffilmark{1}}
\affil{School of Natural Sciences, Institute for Advanced Study, Princeton, NJ 08540}
\affil{gaudi@sns.ias.edu}

\altaffiltext{1}{Hubble fellow}

\begin{abstract}

Recently Sahu et al., using the {\it Hubble Space Telescope} to
monitor stars in the direction of the old ($\sim 12~{\rm Gyr}$)
globular cluster M22, detected six events in which otherwise constant
stars brightened by $\sim 50\%$ during a time of $\la 1~{\rm day}$.
They tentatively interpret these unresolved events as due to
microlensing of background bulge stars by free-floating planets in
M22.  Using simple analytic arguments, I show that if these spike
events are due to microlensing, the lensing objects are unlikely to be
associated with M22, and unlikely to be part of a smoothly distributed
Galactic population.  Thus either there happens to be a massive, dark
cluster of planets along our line-of-sight to M22, or the spike events
are not due to microlensing.  The lensing planets cannot be bound to
stars in the core of M22: if they were closer than $\sim 8~\au$, the
lensing influence of the parent star would have been detectable.
Moreover, in the core of M22, all planets with separations $\ga 1~\au$
would have been ionized by random stellar encounters.  Most unbound
planets would have escaped the core via evaporation which
preferentially affects such low-mass objects.  Bound or free-floating
planets can exist in the outer halo of M22; however, for reasonable
assumptions, the maximum optical depth to such a population falls
short of the observed optical depth, $\tau \sim 3 \times 10^{-6}$, by
a factor of 5-10.  Therefore, if real, these events represent the
detection of a significant free-floating Galactic planet population.
The optical depth to these planets is comparable to and mutually
exclusive from the optical depth to resolved events measured by
microlensing survey collaborations toward the bulge, and thus implies
a similar additional mass of lensing objects.  If the spatial and
kinematic distributions of the two populations are the same, there are
$>10^3$ planets per bulge microlens.  Such a population is difficult
to reconcile with both theory and observations.

\end{abstract}


\section{Introduction}\

Our understanding of the low-mass end of the compact object mass
function has seen considerable progress on many fronts in the last decade.  
The enormous success of radial velocity searches for
extrasolar planets has improved considerably our knowledge of the
statistics, distribution, and mass function of close ($a\la 3~\au$) and
relatively massive ($M\ga 0.2~{\mjup}$) companions to nearby stars
(Marcy, Cochran, \& Mayor 2000; Jorissen, Mayor, \& Udry 2001; Tabachnik \& Tremaine 2001).
Direct searches for brown dwarf companions to local solar-type stars 
have led to only one detection,
GL 229B (Nakajima et~al.\ 1995, Oppenheimer et~al.\ 2001).  However,
serendipitous discoveries in wide-field surveys of very low mass companions to normal stars
(Kirkpatrick et~al.\ 2001) has led to the
conclusion that the `brown dwarf desert,' the paucity of close brown
dwarf companions to stars monitored in radial velocity surveys (Marcy \&
Butler 2000), does not exist for wide separations (Gizis et~al.\
2001).  The statistics of very low mass companions $M \la
10~M_\oplus$, will likely have to wait for future astrometric
(Lattanzi et al.\ 2000) or transit-detecting (Borucki et~al.\ 1997) satellites.

These studies have focused on GKM dwarfs in the immediate local
neighborhood.  Detailed studies of the parent stars of extrasolar
planets have revealed that these hosts have higher metallicity in
comparison to an unbiased field sample (Gonzalez 1997; Laughlin 2000).  
However, it is difficult to
interpret this observation, as it is not clear if the cause of this
enhanced metallicity is stellar pollution from cannibalized planets or
rather that low metallicity tends to prohibit formation
(Murray et al.\ 2001; Santos, Israelian, \& Mayor 2001;
Pinsonneault, DePoy, \& Coffee, 2001).  In order to resolve this
issue, it may be necessary to look toward other systems, such as
globular clusters, open clusters, and the Galactic bulge.  Such
systems are useful in that they act as a control samples, where due to
the homogeneous nature of the systems, one particular parameter that may
affect planet formation and/or evolution (i.e.\ metallicity, age,
local density) can be better isolated.  One of the most promising
methods of studying companions to stars in such systems is transits,
which is in a sense ideally suited to this application since it works
best when a large number of stars can be monitored simultaneously.  An
important null result was found when this tactic was combined with the
high resolution of the {\it Hubble Space Telescope} (HST) to search
for planetary companions of stars in the (relatively) metal-rich globular cluster 47 Tuc.
Gilliland et~al.\ (2001) found that the frequency of close planetary 
companions to stars in 47 Tuc is more than an order of magnitude
smaller than in the local solar neighborhood.  This lack of planets is difficult
to explain in terms of dynamical stripping (Davies \& Sigurdsson
2001) but may be explicable by invoking disk photoevaporation (Armitage 2000).  Observations in other
environments, such as the Galactic bulge (Gaudi 2000) or open
clusters, have been considered, are currently being undertaken, or are being planned.

Large scale optical and infrared surveys such as the {\it Sloan
Digital Sky Survey (SDSS)} and the {\it Two Micron All Sky Survey
(2MASS)} have discovered a significant population of low mass 
objects.  By now nearly 100 L-dwarfs (Fan et~al.\ 2000, Kirkpatrick
et~al.\ 2000), and a dozen T-dwarfs (Strauss et~al.\ 1999, Burgasser
et~al.\ 1999) are known.  The difficulty with using these
observations to constrain the low mass field object mass function is
that for a given effective temperature, there exists a degeneracy
between mass and age, making the determination of the mass function
fundamentally uncertain (Reid et al. 1999).

Observations of low mass field objects confined to stellar aggregations
alleviate some of these difficulties, because the age of the system
is known or can be estimated in many cases.  Such studies have led to
the detection of a population of low-mass, free-floating objects in
several open clusters (Bouvier et al.\ 1998; Hambly et~al.\ 1999;
Zapatero Osorio et~al.\ 2000; Mart{\' i}n et~al.\ 2000).  Many of
these are clearly brown dwarfs, but some are suggested to have
masses that place them in the canonical planet regime ($M<13\mjup$).  The
relation of these objects to those found orbiting local stars is
unknown, and even their designation as planets is
controversial (McCaughrean et~al.\ 2001).

Microlensing is a unique method of detecting planets that offers the
significant advantage that its effect is sensitive only to the mass of the
lensing object, and does not rely on either the flux of the
planet or its parent star.  Therefore distant, extremely faint, or
even completely non-luminous compact objects, either free-floating or
bound to other objects, can be detected using microlensing.  Indeed
microlensing was originally suggested by Pacy{\'n}ski (1986) as a
method to look for dark matter in the halo of our Galaxy.  In
principle, extremely low-mass objects can be detected, with the
ultimate lower limit set by the size of the source stars.  This fact
was exploited by the MACHO and EROS collaborations to rule out objects
having masses of $10^{-7}-10^{-3}\msun$ as the primary constituents of
halo dark matter (Alcock et~al.\ 1998).  The PLANET collaboration
acquired and analyzed five years of photometric data searching for
planetary companions to Galactic bulge stars, using a method first
suggested by Mao \& Paczy\'nski (1991), but found no candidates,
implying that $\sim \mjup$ mass planets with separations of $1.5~\au
\le a \le 4~\au$ occur in less than 1/3 of systems (Albrow et~al.\
2001, Gaudi et~al.\ 2002).  Free-floating planets are also detectable in microlensing
searches toward the Galactic bulge (di Stefano \& Scalzo 1999a).  The
primary drawback to using microlensing to detect low-mass objects is
that the mass of the lensing object typically cannot be directly
measured.  Instead, one measures the timescale $\te$, which is a
degenerate combination of the mass of the lens, and
the lens-source relative parallax and proper motion.  
This makes unambiguous detection of low-mass objects difficult.

Recently, Sahu et~al.\ (2001, hereafter SM22) presented an
ingenious method of overcoming this difficulty, based on a suggestion
originally made by Paczy{\'n}ski (1994).  They used HST to monitor
stars toward the old, metal-poor globular cluster M22 ($l=9\deg.9, b=-7\deg.6$).  Because M22
is projected in front of the bulge, there exists a significant
probability (optical depth) that background bulge stars will be
microlensed by foreground M22 objects.  Since the distance, velocity
dispersion, and proper motion of M22 are known, the detection of a
microlensing event, and a measurement of $\te$ yields a measurement of
the mass of the lens, with only the unknown distance and 
velocity of the source contributing significantly to the error.  In
fact, SM22 did detect one event with $\te=17.6~\days$, yielding a lens
mass of $M= 0.13^{+0.03}_{-0.02}\msun$ -- an impressive mass
measurement of an M-dwarf 2.6~kpc away.

SM22 also detected six events in which a constant light curve
brightened by $\sim 50\%$ during one set of
two measurements, separated by six minutes, and then returned to
baseline for the remaining measurements.  By direct inspection of the
images, SM22 rule out detector artifacts or cosmic rays as the source
of these brightenings.  Furthermore, the
brightening was consistent in both images in each of the six cases.  SM22
rule out several other astrophysical sources of variability, and
tentatively conclude that these brightenings are due to unresolved
microlensing events.  With the sampling rate of $\sim 1$ per $4$ days,
this implies an upper limit to the timescale of $\te \la 1~{\rm day}$.
If the lenses are located at the distance of M22 [$\dol=2.6\pm
0.3~\kpc$ (Peterson \& Cudworth 1994)] and have kinematics similar to
M22 stars, then this translates to an upper limit on the mass of the lenses
of $M_p\la 0.25\mjup$.  SM22 tentatively conclude that they have found
evidence for a considerable population of free-floating planets in
M22.  This is interesting: if every star in M22 had $n_p$ planets
of mass $M_p$ associated with it, than the optical depth $\tau_p$ to
lensing by these planets would be roughly given by
\be
\tau_p \approx {M_p \over M_*} n_p \tau_*,
\label{eqn:taup}
\ee where $M_*$ is the mass of a typical star in M22, and $\tau_*$ is
the lensing optical depth to normal stars. Therefore for $M_p\le0.25\mjup$ and
$\tau_* \simeq 10^{-5}$ for the core of M22 (see \S 5), $\tau_p \la
10^{-8} n_p$.  Since SM22 monitored $\sim8 \times 10^5$ stars,
detecting any events would be very unlikely, unless $n_p$ is very
large.  Indeed, SM22 estimate that 10\% of the cluster mass may be in
these planetary objects.

de la Fuente Marcos \& de la Fuente Marcos (2001) discuss some of
the implications of this inference.  By analyzing a suite of
 N-body simulations, they
argue that the majority of planets in the core of M22 will be ionized from 
their parent stars, and those ionized planets will quickly evaporate
from the core.  They argue that bound planets may be able to survive
in the outskirts of M22, thus reproducing the observed optical depth, but
only if multiplanet systems are common.  However, they favor 
the possibility that these events are due to a dark cluster 
of planets not associated with the globular cluster.

Because microlensing events do not repeat, the only way to directly
verify these observations is to perform the same experiment at even
higher time resolution.  Since this experiment essentially requires
the resolution of HST, this would be a major investment of limited
resources.  Given the importance of this result on the one hand, and
the considerable expenditure of resources required for direct
verification on the other, it seems prudent to consider the SM22
interpretation of these spike events in more detail.  Here I present
such a study, which is complementary to that presented by de la
Fuente Marcos \& de la Fuente Marcos (2001) in that I rely primarily
on analytic arguments (rather than on the results of N-body
simulations). 
In \S 2, I briefly review the SM22 observations, and
collect some relevant parameters for M22. In \S 3, I review the basics
of microlensing, and apply these to the M22 observations in \S 3.1.
In \S 4, I show that these objects are unlikely to be associated with
M22, by first demonstrating that these planets must be separated by $a
\ga 8\au$ from their parent stars (\S 4.1), by second showing that
planets in the core with separations $a \ga 0.3\au$ would be ionized
by other stars (\S 4.2), by third showing that most low mass,
free-floating planets would have evaporated from the cluster core 
over the cluster lifetime (\S 4.3), and finally by showing that the 
maximum optical depth to planets in
the halo of M22 falls considerably short of the observed optical depth
(\S 4.4).  Thus if the spike events are indeed due to microlensing,
the lensing objects cannot be associated with M22.  In \S 5, I explore
the implications of this conclusion, and show that, if smoothly
distributed, the implied
free-floating Galactic planet population is hard to reconcile with
current theoretical and observational constraints.  Thus either
the explanation of de la Fuente Marcos \& de la Fuente Marcos (2001)
is correct -- there is a dark cluster of planets along our line-of-sight
to M22 -- or the majority of the spike events cannot be due to
microlensing. In light of the circumstantial evidence against 
the microlensing interpretation of these events, 
in \S 6 I present a critical re-evaluation of the direct evidence 
in favor of microlensing.  I summarize and conclude in \S 7.

\begin{table*}[t]
\begin{center}
\begin{tabular}{|c||c|c|c|c|c|c|}
\tableline 
Source  & $m_{F606W}$ & $m_{F814W}$ & $\Delta m_{F814}$ & $A$    & $(m_{F606W}-m_{F814W})_0$\tablenotemark{1} & $M_{F814W}$\tablenotemark{2}\\
\tableline
Fig.\ 1 & 18.86     & 17.85     &   ---           & ---  & 0.70     & 2.70 \\
A	& 23.32	    & 21.64     &  -0.46          & 1.53 & 1.37      & 6.49\\
B  	& 23.81     & 22.25	&  -0.46	  & 1.53 & 1.25	     & 7.10\\
C 	& 23.51	    & 22.12	&  -0.72	  & 1.94 & 1.08      & 6.97\\
D	& 22.70     & 21.33	&  -0.31 	  & 1.33 & 1.06      & 6.18\\
E  	& 23.03	    & 21.49	&  -0.49	  & 1.57 & 1.23      & 6.34\\
F	& 22.14	    & 20.81	&  -0.31	  & 1.33 & 1.02      & 5.66\\
\tableline
\end{tabular}
\end{center}
\tablenum{1} {\bf Table 1} Measured and estimated parameters for the M22 events.\\
\tablenotetext{1} {Dereddened color assuming $E(B-V)=0.634$, or $E(F606W-F814W)=0.31$.}
\tablenotetext{2} {Absolute magnitude assuming a distance modulus of 14.52 
and $E(B-V)=0.634$.}
\label{tbl:table1}
\end{table*}

\begin{table*}[t]
\begin{center}
\begin{tabular}{|c|c|c|c|}
\tableline 
Parameter & Symbol & Value & Reference\\
\tableline
Distance  & $\dol$   & $2.6~\kpc$ & Peterson \& Cudworth (1994) \\
Core Radius & $r_c$    & $1'\hskip-2pt.4$  	    & Trager, King, \& Djorgovski (1995)\\
Tidal Radius & $r_t$   & $28'\hskip-2pt.9$      & Trager, King, \& Djorgovski (1995)\\
Central Density & $\rho_0$ & $4.65\times10^3\msun~{\rm pc^{-3}}$ & Peterson \& King (1975)\\
Velocity Dispersion & $\sigma$ & $11.4~\kms$ & Peterson \& Cudworth (1994)\\
Proper Motion & $\mu_l$ & $10.9~{\rm mas~yr^{-1}}$ & Peterson \& Cudworth (1994)\\
Age   & $T_0$  &  12 Gyr & Davidge \& Harris (1996)\\
Typical Mass & $M_*$ & $0.33~\msun$ & Paresce \& De Marchi (2000) \\
Reddening & $E(B-V)$ & 0.326 & Schlegel, Finkbeiner, \& Davis (1998)\\
Galactic Coordinates & $(l,b)$  & $(9\deg.9,-7\deg.6)$ & -- \\
Metallicity & [Fe/H] & $-1.54 \pm 0.11$ & Lehnert, Bell, \& Cohen (1991)\\

\tableline
\end{tabular}
\end{center}
\tablenum{1} {\bf Table 2} Collected relevant parameters for M22.\\
\label{tbl:table2}
\end{table*}

\section{Observations and M22 Parameters}

SM22 monitored the central region of M22 with HST over the course of
114 days, with a total of 43 epochs with three fields monitored at
each epoch.  The majority of the measurements was taken in the F814W
($I$) filter with every fourth measurement in the F606W (wide $V$)
filter.  I will be primarily considering the light curves presented in
Figure 2 of SM22, which show the F814W band light curves of the six spike
events over a period of 105 days, with $\sim 25$ measurements in each
light curve, for a sampling interval of $\sim 4~{\rm days}$.  The
typical photometric errors for these light curves, as judged by the
scatter, is $1-5\%$.  Table 1 gives the the baseline F814W and F606W
magnitudes of these six spike events, reproduced from Table 1 of SM22,
and retaining their lettering scheme.  Also presented in Table 1 are
estimates of the amount by which these events brightened, both in
magnitudes, $\Delta m_{F814W}$ and the corresponding magnification $A$
assuming no blending\footnote{For typical ground-based resolutions
($\sim 1''$), a large fraction of sources are blended, i.e., there
exist multiple stars in each resolution element (Han 1997a).  However,
with the resolution of HST, essentially all stars are resolved (Han
1997b).  Thus the observed flux is most likely directly related to the
magnification, unless the lens itself, or an unlensed companion to the
source or lens is of comparable brightness to the lensed source.}.  Values for $\Delta m_{F814}$ have errors of
$2-7\%$, as estimated by eye.  Finally, Table 1 shows the
dereddened color and absolute magnitude of the SM22 sources, assuming they are located at
the Galactic center, i.e.\ a distance modulus of 14.52 ($8~\kpc$), and
$E(B-V)=0.326$.  The latter was found using the reddening maps of Schlegel,
Finkbeiner, \& Davis (1998).  This value agrees well with that
reported by Piotto \& Zoccali (1999) from main sequence fitting to
M22, and translates to $A_{F606W}=0.94$ and $A_{F814W}=0.63$.

Table 2 lists parameters of M22 collected from various sources in the
literature that will be relevant to the discussion.  In some
cases, there are discrepancies in the published literature.  
Where possible, I have adopted the most modern
determinations.  Adopting other values for these parameters will
change some of the resulting computations in detail, but will not
affect the general arguments and conclusions.

\section{Microlensing Basics}

In this section I review the fundamentals of microlensing, concentrating on 
those concepts that are important for the present study.  A lens of 
mass $M$ at a distance $\dol$ has an angular Einstein ring radius of 
\be
\thetae=\sqrt{{{4 G}\over c^2} {M \over \drel}}\simeq 0.834~{\rm mas}\left({M \over 
0.33~\msun}\right)^{1/2}, 
\label{eqn:thetae}
\ee
where  $\drel$ is defined by,
\be
{1 \over \drel} \equiv {1 \over \dol} - {1 \over \dos},
\label{eqn:pirel}
\ee
and $\dos$ is the distance to the source.  The scaling relation at the
extreme right hand side of \eq{eqn:thetae}, and all scaling relations
presented in this section, are appropriate for M22 parameters, i.e.\
$\dol=2.6~\kpc$, $\dol=8.0~\kpc$.
At the distance of the lens, $\thetae$ corresponds to a physical
distance of 
\be
\re = \thetae \dol \simeq 2.17~\au \left({M \over 
0.33~\msun}\right)^{1/2}.
\label{eqn:re}
\ee
The timescale of a microlensing event is given by,
\be
\te={\thetae \over \murel}\simeq 27.9~\days \left({M\over
0.33~M_{\odot}}\right)^{1/2},
\label{eqn:te}
\ee
where $\murel$ is the lens-source relative proper motion. 
I will assume that the mean source proper motion is zero (which is appropriate
for sources in the bulge), and
thus $\murel=\mu_l=10.9~{\rm mas~yr^{-1}}$.

The foreground lens magnifies the background source
by an amount that depends only on the angular separation $\theta$  between the
lens and source in units of $\thetae$, $u\equiv \theta/\theta_E$.
For a single lens the magnification is,
\be
A(u)= {{u^2+2} \over {u \sqrt{u^2+4}}}.
\label{eqn:mag}
\ee
In particular $A(1)=3/\sqrt{5}\simeq 1.34$. For a simple microlensing,
\be
u(t)  = \left[u_0^2+ \te^{-2}(t-t_0)^2\right]^{1/2},
\label{eqn:uoft}
\ee
where $t_0$ is the time of maximum magnification and $u_0$ is the impact parameter.  Note that $u_0$ is distributed uniformly. 

The optical depth to microlensing, $\tau$, defined as the probability
that any star is magnified by $>1.34$, is given by,
\be
\tau= {4 \pi G \over c^2}  \int_0^{\dos} d\dol \rho(\dol) \drel
\label{eqn:optdepth}
\ee
where 
$\rho$ is the mass density along the line of sight to the source.  

Observationally, the optical depth can be estimated from a sample of
$N_e$ events with measured timescales $t_{{\rm E},i}$ by 
\be
\tau_{obs}={\pi \over {2 N_* \dt}}\sum_{i=1}^{N_e} {t_{{\rm E},i}\over
\epsilon_i}, \qquad {\rm (Resolved~Events)}
\label{eqn:tauobs}
\ee 
where $N_*$ is the number of stars monitored for a duration $\dt$,
and $\epsilon_i$ is the detection efficiency of event $i$.  As I is
show in \S 3.1, the expression for unresolved events (where $\te$ is
not known) takes a slightly simpler form.

\subsection{Application to the M22 Events}

Before discussing constraints on the location of the lenses, I first
apply the microlensing formalism just presented to the M22 events.
SM22 reports that the single resolved event has a timescale of
$\te=17.6~{\rm days}$.  Assuming that the lens is associated with M22,
and that the source is in the bulge, this translates into a mass of
$M=0.13\msun$ [\Eq{eqn:te}].  Careful accounting of the dispersion in
bulge source distances and kinematics establishes an error of $\sim
15\%$ (SM22).  Inspecting the mass function (MF) of M22 as determined by
Paresce \& De Marchi (2000), the number of stars with this mass is
approximately 20\% smaller than at the peak of the MF ($M=0.33\msun$).
One would expect $\sim 0.8 (0.13/0.33)^{1/2}\sim 50\%$ fewer events
from objects with $M=0.13\msun$ than the peak of the MF, and thus the
derived mass is quite plausible. The optical depth implied by this one
resolved event ultimately depends on the detection efficiency, but a
lower limit can be found by assuming $\epsilon=1$.  Due to the high
quality of the HST photometry, the fact that the sampling interval is
considerably less than $\te$, and that blending is likely not an issue, the
true optical depth is probably not much larger than this lower
limit. I find, \be
\tau_{res} \ga 3 \times 10^{-6}.
\label{eqn:taures}
\ee
Where I have adopted $N_*=83,000$, as indicated by SM22, and
$\dt=105~{\rm days}$, which is derived from their Figures 1 and 2.  As
I show in \S 4.4, this lower limit is a factor of $\sim 3$ smaller than
would be expected based on what we think we know about the central
density and structure of M22.  This could be caused by inaccurate
parameters for M22, an efficiency considerably smaller than unity, or
Poisson fluctuations.  The one resolved event is somewhat unusual
because the source is reported to be a variable.  In fact, in order to
derive $\te$ from the light curve, SM22 fit the event to a
binary-source model.  This provides a consistency check on the
microlensing interpretation, as the binary-source model provides a
limit on $\thetae$ which can be compared with the value derived from
the timescale (Han \& Gould 1997, Alcock et~al 2001).
  
Obviously the timescales of the six unresolved events are unknown.  
SM22 assert that the statistical upper limit to $\te$ is
$0.8~{\days}$ (at a 95\% confidence level), which translates to $M\le 0.27\times 10^{-4}\msun$ or
$M\la 0.25 \mjup$ where $\mjup$ is the mass of Jupiter.  
A lower limit to the mass can be derived by considering finite
source effects, and is $M \ga 3\times 10^{-8}$, or about the
mass of the Moon.  The reason that this limit is not very restrictive
is that the source stars are K and M-dwarfs, which are relatively small.  Although the
timescales of the events are unknown, the optical depth can still be
estimated.  In the limit of unresolved (``spike'') events, the optical
depth becomes, 
\be 
\tau_{obs} = {N_{\rm spike} \over {N_* N_d}} {1 \over {\min(1,\ut^{2})}}, \qquad {\rm (Spike~Events)}
\label{eqn:tauspike}
\ee
where $N_{\rm spike}$ is the number of spike events having $A>1.34$, $N_d$ is the number of 
epochs per light curve, and $\ut$ is,
\be
\ut=\sqrt{2}\left[(1-\at^{-2})^{-1/2}-1\right]^{1/2},
\label{eqn:uthresh}
\ee
and $\at$ is the minimum detectable magnification.  In the case of the 
six spike events, $\ut > 1$, and thus the optical depth
is simply $\tau_{obs}=N_{\rm spike} N_*^{-1} N_d^{-1}$.  Adopting $N_d=25$,
\be
\tau_{\rm spike}= (2.9 \pm 1.2) \times 10^{-6}
\label{eqn:tauobsspike}
\ee
where the error is that due solely to Poisson statistics.  
The optical depth to the single resolved event can also be
estimated using \eq{eqn:tauspike}, simply by setting $N_{\rm spike}$ equal
to the number of data points on the event light curve with magnification $> 1.34$, restricting attention to those taken in the F814W filter. Since 
there are six such points, this results in the same optical depth
as $\tau_{\rm spike}$, in excellent agreement with the optical depth
estimated from the usual formula.    
Thus the optical depth due to the spike events is comparable to that
contributed by the one resolved event.  If the spatial and kinematic
properties of the two samples are the same, this implies a similar
mass in each component.

\section{Binary Lenses, Ionization, and Evaporation}

\subsection{Binary Lenses}

All six of the spike events discovered by SM22 are well characterized
by a flat light curve (to within the photometric errors) with one
deviant data point.  As pointed out by SM22, the fact that the light
curve exhibits no features other than the one deviant point implies
that the planet must be quite distant from any parent star, or the
lensing influence of the primary would be detectable.  The minimum
separation compatible with the data depends on the mass of the primary
and the photometric accuracy, $\sigma_P$.  Formally, the lensing
behavior of the planet bound to a parent star is described by the
formalism of binary lenses (Mao \& Paczynski 1991; Gould \& Loeb
1992).  However, when the projected separation of the planet and star
separation in units of $\thetae$ is much larger than the Einstein ring
radius of the system, the magnification structure is well described by
the superposition of two point masses
(di Stefano \& Mao 1996, di Stefano \& Scalzo 1999b)
separated by a distance of $\dtil=(d^2-1)/d$, 
where $d$ is the instantaneous angular separation of the planet
and star in units of $\thetae$ (Gaudi \& Gould 1997).  The 
detection probability is roughly the probability that the
source trajectory will pass within $\ut$ of the primary
lens multiplied by the probability that the deviation due to the primary
occurs during the observation window.  Assuming that the mass of the
secondary is much smaller than the mass of the primary, that the times of
maximum magnification of the planetary events are uniformly
distributed in the observation window, and normalizing all distances
to the mass
of the primary, the detection probability is,
\be
\pprim(d) =\cases{ 1 & if $\dtil \le \ut$, \cr
{2 \over \pi}\left[1-{\dtil-\ut\over \Delta u}\right]{\rm asin}\left({\ut\over\dtil}\right) 
  & if $\ut<\dtil< \ut + \Delta u$, \cr
0 & if $\dtil \ge \ut+\Delta u$, \cr
}
\label{eqn:pprim}
\ee
and $\Delta u \equiv \Delta t /\te$, and $\te$ is the timescale of the
primary.  In order to convert from the detection probability as a
function of $d$ to the detection probability as a function of the
semi-major axis $a$ of the planet, it is necessary to convolve
$\pprim(d)$ with the probability of $d$ given $\ap \equiv a/\re$,
\be
\pprim(\ap)= \int_0^{\ap} {\rm d}d~\pprim(d) P(d;\ap),
\label{eqn:pconv}
\ee
where 
\be
P(d;\ap)= { d \over \ap}\left( 1- { d^2 \over \ap^2}\right)^{1/2},
\label{eqn:distfunc}
\ee
(Gould \& Loeb 1992).  $\pprim(\ap)$ can be converted to physical units
by adopting a value for $\re$.  Figure 1 shows $\pprim (a)$ as a
function of $a$ assuming a primary of mass $M=0.33\msun$ and
$\at=1+\sigma_P$ with $\sigma_P=1\%, 5\%$, and $10\%$.  The 95\%
confidence level (c.l.) lower limit on $a$ is the value where
$\pprim(a)=1-0.05^{1/6}\simeq 40\%$.  I find
\be
a \ge 7.5~{\au}, \qquad (95\%~{\rm c.l.})
\label{eqn:amin}
\ee
for $\sigma_P=10\%$.  Figure 1 also shows the results of computing
$\pprim(\ap)$ adopting the full binary formalism (Witt 1990), which gives nearly
identical results.

The lower limit given in \eq{eqn:amin} is somewhat model dependent;
adopting a smaller mass for the primary would lower the limit.
However, the estimate is conservative, as I have assumed that the
magnification from the primary must rise above $\at$ for it to be
detected, when in fact, the cumulative effect of the curvature due the
primary on the light curve may make it detectable for separations
considerably larger than those calculated using \eq{eqn:pprim}.

Note that de la Fuente Marcos and de la Fuente Marcos (2001) give a
lower limit on $a$ that is a factor of $\sim 2$ smaller than
\eq{eqn:amin}.  This is due to the fact that they adopt a rather
different -- and considerably more conservative -- detection criterion
than assumed here.  They require that $d \le \ut$ for the primary to
be detected.  This requirement gives the absolute lower limit on $d$
in the sense that the primary will be detectable in essentially all trajectories
if $d \le \ut$.  However, for larger $d$, a fraction $\pprim(d)$ of trajectories
should show evidence for a primary.  Thus the lack of detections
in six events can be used to place a more stringent lower limit.

\subsection{Ionization}

In dense stellar environments, such as the core of M22, planetary
systems cannot be treated as isolated.  The large number density and
velocity dispersion imply that a star is likely, within the lifetime
of the cluster, to encounter another star.  Such encounters will
inevitably lead to stripping off of weakly bound outer planets, i.e.\
ionization.

The average number of encounters a
given planetary systems has with a field star per unit time is given
by $1/\tenc$, where $\tenc$ is the encounter time (Binney \& Tremaine 1997),
\be
\tenc =\left[16 \sqrt{\pi} \nu \sigma a^2 \left( 1+ {GM_* \over
2 \sigma^2 a}\right)\right]^{-1},
\label{eqn:tenc}
\ee
where $\nu$ is the local number density. 
For $a=1~\au$, and assuming parameters appropriate for M22, $\tenc=
4.3~{\rm Gyr}\left(\nu/1.41\times 10^{4}~{\rm pc}^{-3}\right)^{-1}$. Thus,
in the core of M22, every star is likely to experience an encounter 
with another star with an impact parameter $\le 1~\au$ during the lifetime
of the cluster.  According to Heggie's Law, soft binaries will tend to get disrupted 
by such encounters.  The binary is soft if $a > a_{\rm h/s}$, where
\be
a_{\rm h/s}= {2 G M_P \over \sigma^2} \simeq 3\times 10^{-3}\au 
\left(M_P \over 0.25\mjup\right) \left(\sigma\over 11.4\kms\right)^{-2},
\label{eqn:ahs}
\ee
and I have assumed the mass of the planet, $M_p$, is much smaller than the field and
parent star masses, which I have assumed to be equal.   Thus essentially all
planetary systems are soft, and will be disrupted on a time scale $\tenc$.
I therefore write the ionization probability as a function of $a$ as,
\be
\pion(a)=1-e^{-T_0/\tenc(a)}.
\label{eqn:pion}
\ee
This probability is shown in Figure 1 for three different values of 
the number density.  Equations (\ref{eqn:tenc}) and (\ref{eqn:pion}),
although crude, agree well with more detailed and realistic calculations
(Smith \& Bonnell 2001; Bonnell et~al.\ 2001; Davies \& Sigurdsson 2001).
For the core of M22 ($\nu=1.41 \times 10^4~{\rm pc}^{-3}$), 
the upper limit to the separation of a bound planet is,
\be
a \le 0.3~\au.\qquad (95\%~{\rm c.l.})
\label{eqn:amax}
\ee
Comparing equations (\ref{eqn:amax}) and (\ref{eqn:amin}), it is clear
that the spike events cannot be due to planets bound to stars in the
core of M22.

There are several caveats.  First is that I have not considered the possibility of 
exchanges due to encounters, i.e.\ where the planet is transferred from the 
original parent star to the field star as a result of the encounter, rather than being ionized 
completely.  The timescale for exchange is (Hut \& Bahcall 1983; Hut 1983)
\be
\tex={3 \sigma^5 a \over 20 \pi \nu G^3 M_*^3 },
\label{eqn:tex}
\ee
where I have again assumed that the mass of the planet is much smaller
than that of the field and parent stars, with the latter two masses to
be equal. For conditions in the core of M22, $\tex = 1.1~{\rm Gyr} (a
/\au) (\nu/1.41\times 10^{4}~{\rm pc}^{-3})^{-1}$. The
ratio of the exchange to encounter timescales is given by,
\be
{\tex \over \tenc}\simeq 0.25 \left({a \over \au}\right)^3,
\label{eqn:exoverenc}
\ee
where I have neglected the term in $\tenc$ due to gravitational
focusing.  Note that ${\tex/\tenc}$ is independent of $\nu$.  For
$a\ga 1.6~\au$, the rate of encounters will exceed the rate of exchanges,
and the net effect of encounters will be to ionize planets completely.
Since this is safely below the lower limit on $a$ set in \S 4.1,
exchanges can be ignored.

The second caveat is that I have not considered processes that may
{\it create} planetary systems, such as three body interactions
(Heggie 1975) and tidal capture (Fabian, Pringle, \& Rees 1975, Press
\& Teukolsky 1977; Lee \& Ostriker 1986).  In fact, it can be shown
that all such processes are extremely subdominant, and the net result
of dynamical interactions on soft binaries is disruption (see
Appendix 8.B of Binney \& Tremaine 1997).

\centerline{{\vbox{\epsfxsize=10.5cm\epsfbox{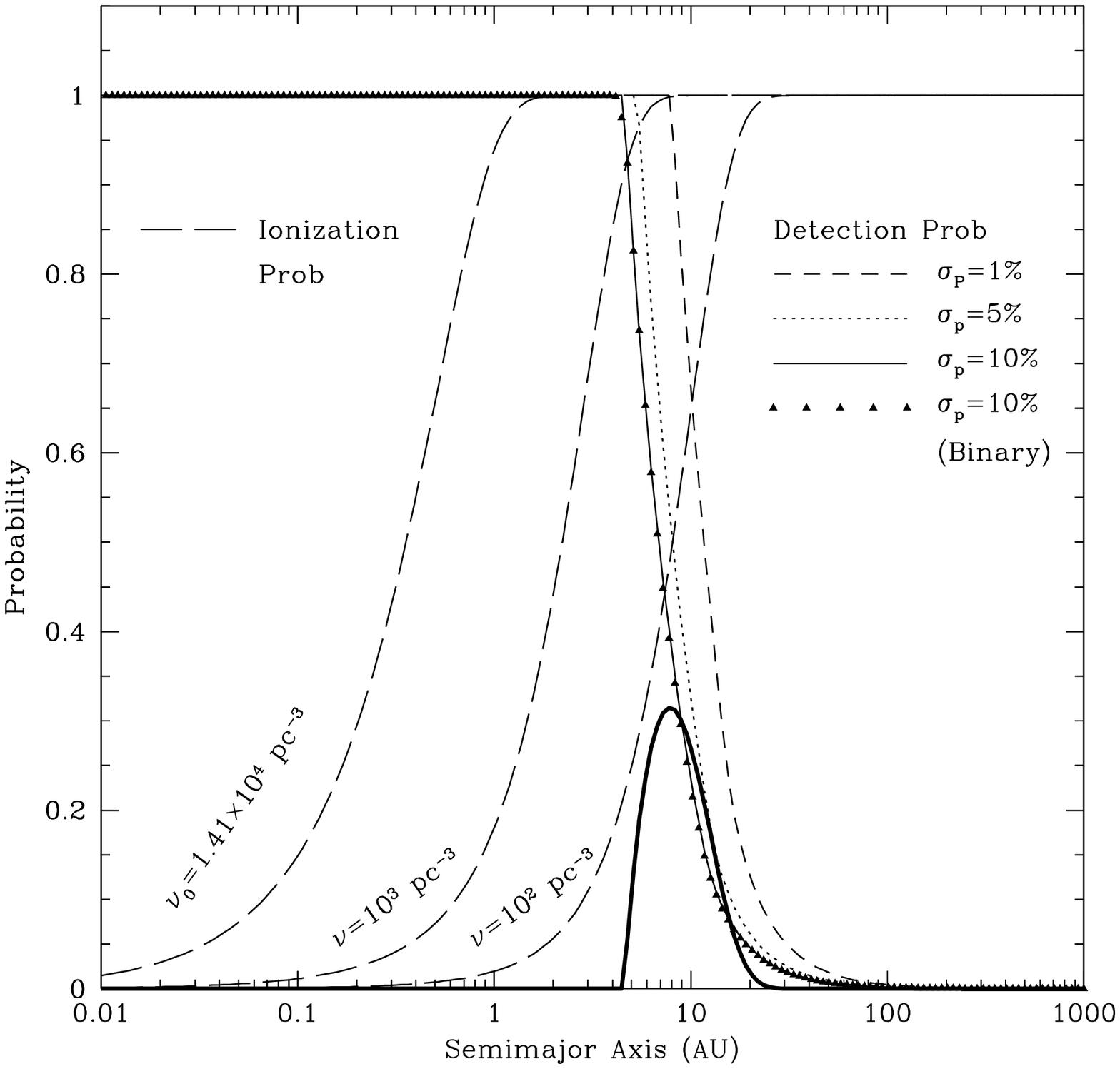}}}}
{\footnotesize {\bf FIG. 1}
The long-dashed curves show the ionization probability $\pion$ as a
function of semi-major axis assuming a cluster age of $T_0=12~{\rm
Gyr}$, a velocity dispersion of $\sigma=11.4~\kms$, and stellar
densities of (left to right) $\nu=1.41\times10^4, 10^3$ and $10^2~{\rm pc^{-3}}$.
The solid, dotted, and short-dashed curves show the primary detection
probability $\pprim$ as a function of the semi-major axis of the
planet for photometric errors of $\sigmap=1\%, 5\%$, and $10\%$,
respectively.  The triangles show the detection probability for
$\sigmap=10\%$ using the full binary-lens formalism.  The bold solid
curve is the probability of not being ionized and not being detected,
i.e.\ $(1-\pion)(1-\pprim)$, for $\sigmap=10\%$ and $\nu=10^2~{\rm
pc^{-3}}$
}
\bigskip

\subsection{Evaporation}

Some fraction of the planets that are ionized from their parent stars
will have, upon ionization, speeds exceeding the local escape speed,
which for the core is $v_e \simeq 23~\kms$ (Peterson \& Cudworth 1994).
These planets will rapidly escape the cluster on a timescale of order
the crossing time, $\tcross=r_t/v_e \simeq 1~{\rm Myr}$.  However,
this will constitute a very small fraction of all ionized planets.
The majority of the ionized planets will remain bound to the cluster
with velocities similar to the velocities of their parent stars.  Over
the lifetime of the cluster, these free-floating planets will undergo
many encounters with the cluster's stellar population.  These encounters
will tend to drive the system toward equipartition, so that
$\sigma_p^2 \sim (M_*/M_p) \sigma_*^2$, where $\sigma_*$ is the velocity
dispersion of the stars.  If the cluster
mass is dominated by the stellar component, i.e.\ the total mass in
planets is negligible, then the potential will be determined by the
stars, and the local velocity dispersion $\sigma$ of the cluster will be $\sigma_*$.
In this case, it is clear that equipartition cannot actually
be realized, since $M_*/M_p > 4$, and hence the planets will attain
velocities larger than the local escape speed, $\sigma_p > 2 \sigma =
v_e$, and thus escape from the system.

The timescale over which the planets get ejected from the
system is the relaxation time (Spitzer 1987),
\be
\trelax =  0.065 {\sigma^3 \over \nu M_*^2 G^2 \ln{\Lambda}},
\label{eqn:trelax}
\ee
where $\ln{\Lambda}$ is the Coulomb logarithm which I will take to be
$\ln{\Lambda}=\ln{(0.4N_*)}=\ln(4\pi\nu_0r_c^3/3)\simeq 10$.  Here $N_*$
is the total number of stars in the system. For the core of M22,
$\trelax=0.33~{\rm Gyr}$.  Thus, assuming a constant density and
velocity dispersion, the core has experienced $\sim 35$ relaxation
times during the lifetime of the cluster.  Therefore all planets
should have long since evaporated from the cluster core.

Note that \eq{eqn:trelax} is the relaxation time for the stars, which
is appropriate for the planets only if the mass in
planets is much smaller than that in stars.  In the case when the two
mass components are comparable, one might be tempted to adopt the
equipartition timescale (Spitzer 1969) 
\be 
\teq = {\sigma^2 \over {4
\sqrt{3 \pi} G^2 M_* M_p \nu \ln{N} }} \simeq {M_* \over M_p} \trelax.
\label{eqn:teq}
\ee 
However, as argued before, the system cannot actually achieve
equipartition, because $M_*/M_p > 4$.   Thus the estimate
in \eq{eqn:teq} breaks down, and the true evaporation timescale will
be considerably smaller.  Regardless, if the mass
in planets were comparable to that in stars today, 
M22 should exhibit an relatively
large mass-to-light ratio, since planets of $M_p\simeq 0.25\mjup$ emit
very little light.  Peterson \& Cudworth (1994) find a global mass-to-light
ratio of $M/L \la 1$ for M22,
which is anomalously low for globular clusters, and not compatible with 
a substantial population of planets by mass.  Therefore \eq{eqn:trelax} should be applicable.

\subsection{Planets in the Halo of M22?}
 
The results of \S\S4.1, 4.2 \& 4.3 show that a significant population
of planets, either bound or free-floating, cannot exist in the core of
M22 (and explain the spike events).  However, the relevant timescales,
$\tenc$ and $\trelax$, are both inversely proportional to the stellar
density $\nu$, which drops precipitously outside the cluster core,
typically as $\nu \propto r^{-3}$, where $r$ is the radial distance
from the center of the cluster.  For the low density outskirts of the
cluster, the encounter and relaxation timescales can be larger than the
lifetime of the cluster, and bound and free floating planets can
survive.

Figure 1 shows $\tenc$ for stellar densities of $\nu=10^3~{\rm
pc^{-3}}$ and $10^2~{\rm pc^{-3}}$.  For very low densities, bound
planets are in principle compatible with the observed spike events.
For example, for $\nu = 10^2~{\rm pc^{-3}}$, there exists a range of
$a$ (albeit a small one) where bound planets could survive ionization
and yet still satisfy the lower limit of $a>7.5~\au$ set by the lack of
detection of the parent star.

Similarly, for the outskirts of the cluster the relaxation time can
exceed the lifetime of the cluster.  Specifically, assuming $\sigma$
is constant throughout the cluster, $\trelax > T_0$ for $\nu \la 5.5
\times 10^2 ~{\rm pc^{-3}}$.  In these regions free-floating planets
will evaporate very slowly.

\centerline{{\vbox{\epsfxsize=10.5cm\epsfbox{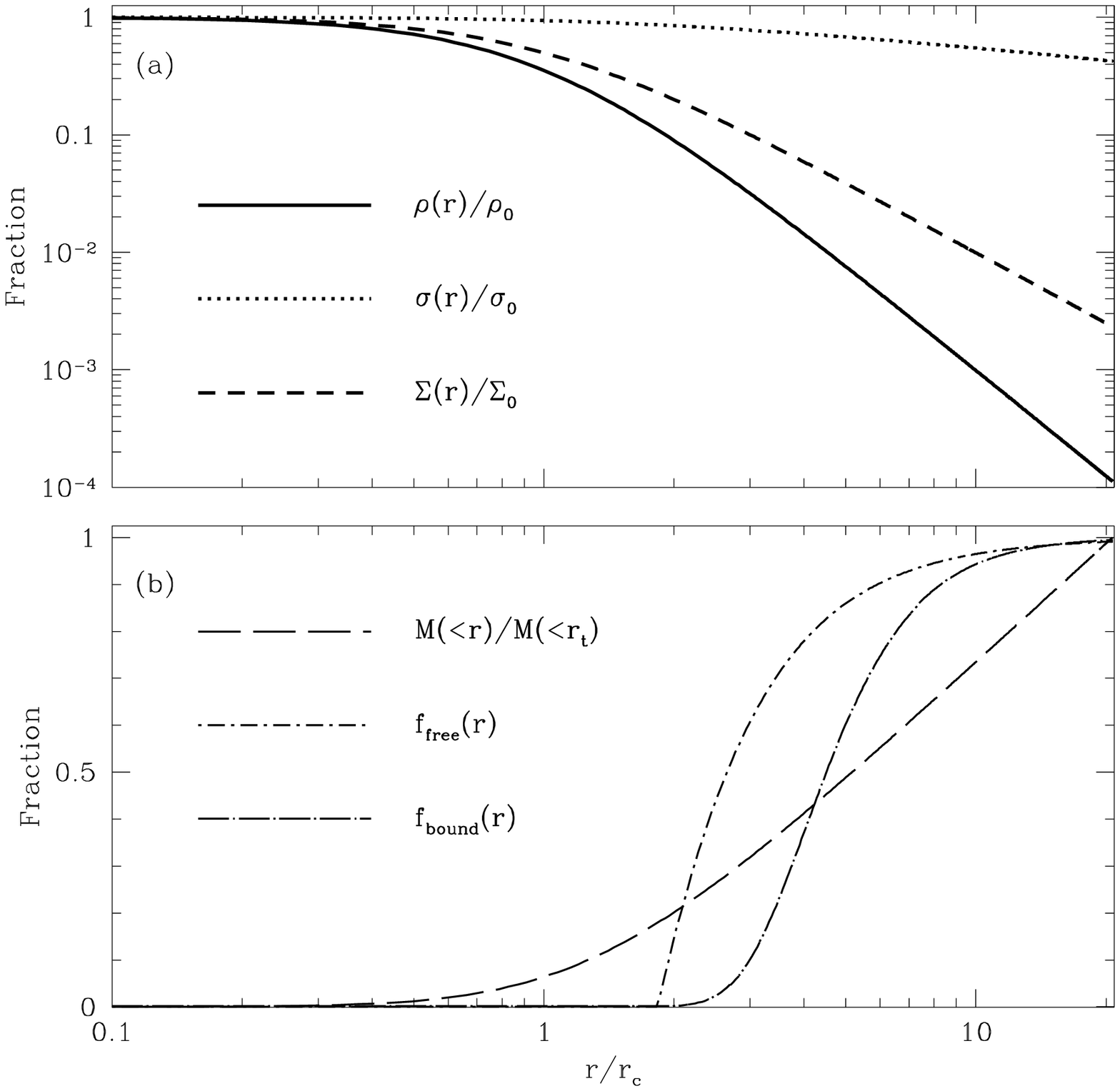}}}}
{\footnotesize {\bf FIG. 2}
(a) The solid curve shows the physical density $\rho(r)$ normalized to the central 
density $\rho_0$ as a function of radius from the center of the cluster,
$r$, normalized to the core radius $r_c$, 
for a model with $\rho(r) =\rho_0 (1+r^2/r_c^2)^{-3/2}$.
For M22, $\rho_0=4.65\times10^3\msun~{\rm pc^{-3}}$, $r_c=1'\hskip-2pt.4$,
and the tidal radius is $r_t=29'\hskip-2pt.0\simeq 21r_c$ (at the right edge of the figure).
The dotted line shows the local velocity dispersion $\sigma$ 
normalized to the central velocity dispersions $\sigma_0$. 
The short-dashed line shows the projected mass density $\Sigma(r)$, where
$r$ now refers to the projected distance from the cluster center.   
(b) The long dashed line in the mass $M(<r)$ interior to $r$, normalized
to $M(<r_t)$, the total mass interior to the tidal radius, $r_t$.
The dot-short-dashed line shows $f_{\rm free}$,
the fraction of free-floating planets that have survived evaporation. 
The dot-long-dashed line shows $f_{\rm bound}$, the fraction of 
systems that remain bound.}
\bigskip

Can the bound or free-floating planets in the halo of M22 explain the
measured optical depth? The answer will depend on the competition
between the declining contribution to the optical depth from the outer
parts of the halo and the increasing longevity of
bound and/or free-floating planets.  Therefore it is necessary to assume a
form for the density profile of M22.
I will adopt the following form,
\begin{equation}
\rho(r) = \rho_0 \left[ 1 + (r /r_c)^2 \right]^{-3/2}\label{eqn:plummerrho}.
\end{equation}
Although a good match to observed cluster properties, this form has the disadvantage
that the total enclosed mass diverges logarithmically.  For practical purposes, I will denote the
total cluster mass as the mass interior to $r_t$.  From this form, the projected mass density, $\Sigma(r)$, as function of the
projected distance from the cluster center can be derived,
\begin{equation}
\Sigma(r)=2\rho_0 r_c \left[1  + (r /r_c)^2 \right]^{-1}.
\label{eqn:plummersur}
\end{equation}
Figure 2 shows $\rho(r)$, $\Sigma(r)$, $\sigma(r)$, and
$M(r)$ as a function of $r/r_c$.  For M22, the concentration is
$c=\log{(r_t/r_c)}=1.31$, and $r_t=20.4r_c$ is just at the edge of the
figure.

From \eq{eqn:optdepth}, and given that size of the cluster is small
compared to $\dos$,  
the optical depth to microlensing is,
\be
\tau= { 4 \pi G \Sigma \dos \over c^2} x(1-x),
\ee
where $x\equiv \dol/\dos$.  Inserting values appropriate for M22, and adopting the form 
for $\Sigma(r)$ given in \eq{eqn:plummersur}, I find,
\be
\tau_{\rm tot} \simeq 10^{-5} [1+ (r/r_c)^2]^{-1}
\label{eqn:taum22}
\ee
Where $r$ refers to the projected distance from the cluster center.
Note that this overestimates the optical depth slightly, since the
form of $\Sigma(r)$ given in \eq{eqn:plummersur} is found from
integrating the density along the line of sight from $-\infty$ to
$\infty$, when in fact the integral should be cut off at the tidal
radius $r_t$.  The difference is negligible except near projected
distances of $r \sim r_t$, where
the optical depth is extremely small anyway.  The optical depth as a
function $r$ in shown in Figure 3.  The HST observations of SM22 were
concentrated in the inner $\sim 2.5'$ of M22.  In this region, $\tau
\simeq 10^{-5}$, which is a factor of $\sim 3$ times larger than the
minimum optical depth estimated from the single resolved event (see \S
3.1).

What fraction of optical depth can be contributed by planets?
Assume some fraction $f_p$ of the cluster mass density was originally
in the form of planets.  For simplicity, I will assume that this
fraction is universal, i.e.\ the primordial mass density in planets
is simply $\rho_p(r)=f_p \rho(r)$.  I will further assume that $f_p$ is
sufficiently small that the dynamics of the cluster is everywhere
dominated by the stellar component, and furthermore that the cluster
has not evolved significantly during its lifetime (almost certainly an
oversimplification).  The fraction of bound planets that will have survived
ionization due to stellar encounters is simply 
\be
f_{\rm bound}(r;a)=1-\pion(r;a)=e^{-T_0/\tenc},
\label{eqn:fion}
\ee
where $P_{ion}$ is given in \eq{eqn:pion}, and the dependence on $r$ arises
because $\tenc$ depends on $\nu$ and $\sigma$.  An upper limit on $f_{\rm bound}(r)$ is found by inserting $a=7.5\au$, which is the lower limit on $a$ set by the lack of detection of the primary in the spike events (\S 4.1).   The fraction of free-floating planet that have survived evaporation from the cluster is,
\be
f_{\rm free}(r)=1-{\xi_e T_0 \over \trelax},
\label{eqn:fevap}
\ee
where $\xi_e$ is the evaporation probability.  Equation
(\ref{eqn:fevap}) implies a linear dependence of the mass loss on
time, as derived analytically for tidally truncated clusters by
Spitzer (1987), and found numerically from the detailed evolutionary
models (Chernoff \& Weinberg 1990; Vesperini \& Heggie 1997).  For a
relaxed system, $\xi_e$ can be estimated as the fraction of stars in a
Maxwellian velocity distribution that have velocities $>v_e=2\sigma$,
which is $\xi_e=7.4 \times 10^{-3}$.  However, this is certainly an
underestimate for the evaporation probability of the planets, because
equipartition will drive them to significantly higher velocities than
the mean $\sigma$.  I adopt the value of $\xi_e=0.156$ for test masses
given in Spitzer (1987), as derived from the models of
H{\' e}non (1961).  Note that this value of $\xi_e$ is the global
probability for the cluster referenced to the
half-mass relaxation time, whereas for this calculation 
an estimate of the local
evaporation probability, referenced to the local relaxation time $\trelax$
is necessary.  I will simply assume the global value of $\xi_e=0.156$
is appropriate, but note that this may be slightly in error. 

\centerline{{\vbox{\epsfxsize=10.5cm\epsfbox{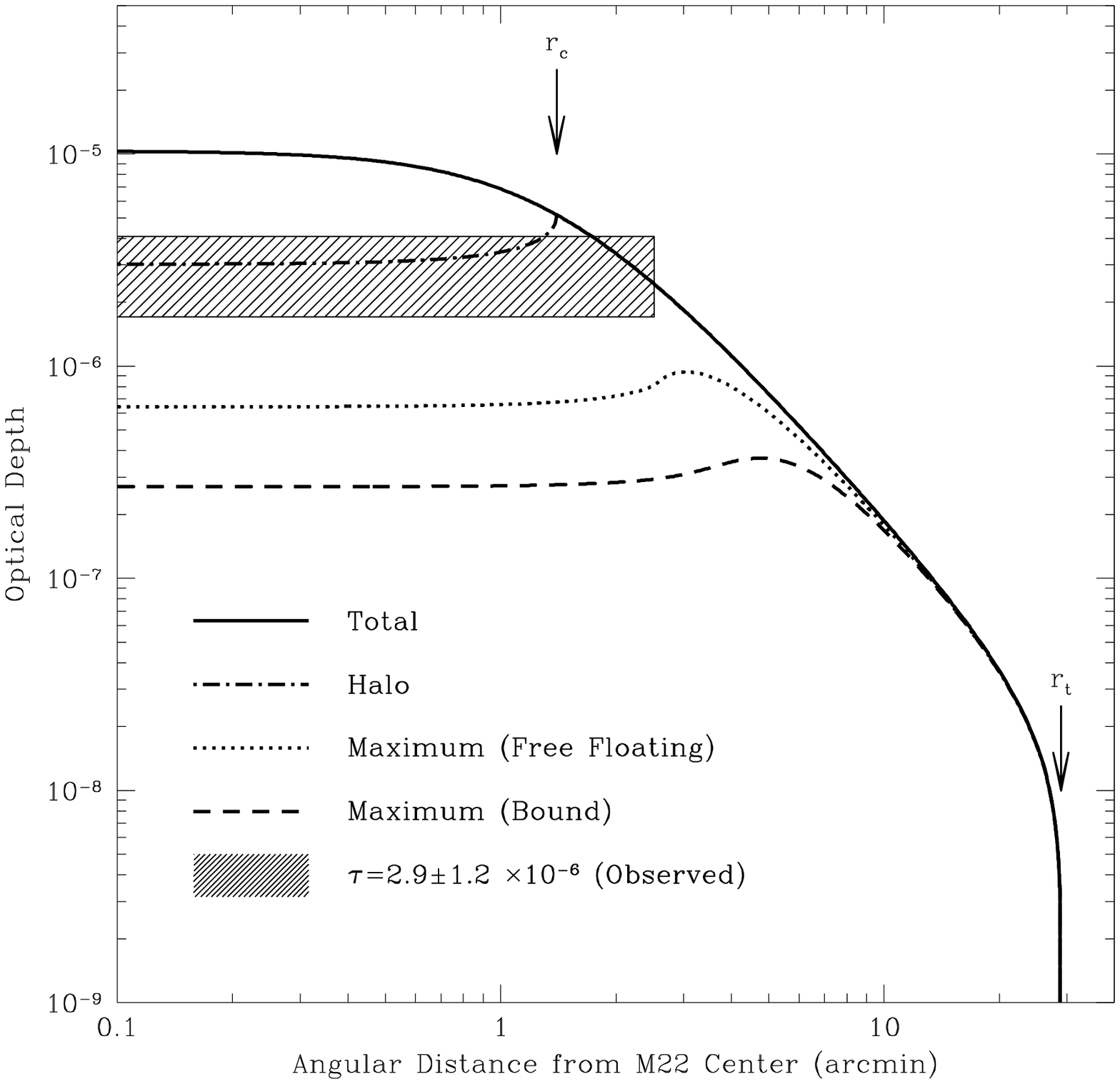}}}}
{\footnotesize {\bf FIG. 3}
The solid line shows the total optical depth $\tau_{\rm tot}$ as function
of angular distance from the center of M22 in arcminutes for the
model shown in Figure 2.  The core radius ($r_c=1'\hskip-2pt.4$) 
and tidal radius ($r_t=28'\hskip-2pt.9$) are marked with
downward-pointing arrows.  The shaded
box shows the inferred optical depth and error from the six spike events, $\tau_{\rm spike}= (2.9 \pm 1.2) \times 10^{-6}$.  The angular extent of the box is the longest width of the WFPC2 camera, $2'\hskip-2pt.5$, and roughly corresponds to the region surveyed by SM22. 
The dotted line is the maximum optical depth from free-floating planets in
such a model.  The dashed line is the maximum optical depth from bound planets.
The dashed-dot line is the halo contribution to the total optical depth.
}
\bigskip

The optical depth contributed from bound planets can now
be estimated,
\be
\tau_{\rm bound}={4 \pi G \dos^2 \over c^2}x(1-x)  
f_p\left[\int {\rm d}s \rho(s) f_{\rm bound}(s)\right],
\label{eqn:maxtaup}
\ee
where the integral is along the line-of-sight
over the extent of the cluster.  The expression
for $\tau_{\rm free}$ is the same, but with $f_{\rm bound}$ replaced by
$f_{\rm free}$.  The value for $f_p$, the fraction of the original
cluster mass in planets, is unknown.  Therefore, I will simply
consider the maximum optical depth, which is found by setting $f_p=1$.
Note that this is truly a limit in the sense that it does not actually
correspond to a self-consistent physical picture.  I discuss this in
more detail below.

The resulting maximum optical depths to bound planets, $\tau^{\rm
max}_{\rm bound}$, and to free-floating planets, $\tau^{\rm max}_{\rm
free}$, are shown in Figure 3.  For observations within the central
$\sim 2'$ of M22, $\tau^{\rm max}_{\rm bound}=6.4 \times 10^{-7}$, and
$\tau^{\rm max}_{\rm free}=2.7 \times 10^{-7}$.  These upper limits
are $\sim 5$ and $\sim 10$ times smaller than the optical depth inferred from
all six spike events, $\tau_{\rm spike}\simeq 3 \times 10^{-6}$, and
are only marginally consistent with the optical depth contributed by just one
event.  Note that such models require that that outer 70\%
of the cluster be {\it entirely} composed of planets.

How sensitive are the results to the adopted form for the density profile?
For isolated clusters, the density should fall off as $\rho(r) \propto r^{-3.5}$ in the outer
halo (Spitzer \& Thuan 1972), whereas tidally truncated clusters will
exhibit somewhat shallower profiles (Chernoff \& Weinberg 1990).  Both
of these profiles are steeper than the $\rho(r) \propto
r^{-3}$ behavior exhibited by the model adopted here.  Indeed, the surface
brightness profile of M22 falls off as $\mu(r) \propto r^{-2.2}$ in
the outer halo (Trager et~al.\ 1995), somewhat steeper than the
$\Sigma \propto r^{-2}$ expectation of the adopted model assuming mass traces light.
In fact, the maximum optical depths for free-floating and bound
planets are not very sensitive to the density profile.  To
first order, the fractions $f_{\rm bound}$ and $f_{\rm free}$ are
inversely proportional to $\rho(r)$.  Therefore, to first order, the
dependence of the maximum optical depth on the density structure
cancels out, and only the dependence of the velocity dispersion
$\sigma(r)$ on the radius enters into the calculation of the maximum
optical depth.  Since this is a relatively insensitive function of the
density structure, the maximum optical depths calculated previously
are reasonably robust to variations in the space of realistic density
models.  I have also computed the optical depth for a Plummer
model which has a density structure $\rho(r) \propto
[1+(r/r_c)^2]^{-5/2}$.  This yields a total central optical depth
almost identical to that for the fiducial model, $\tau\simeq 10^{-5}$.
The maximum optical depths are $\tau^{\rm max}_{\rm bound}=2.8 \times
10^{-7}$, and $\tau^{\rm max}_{\rm free}=1.4 \times 10^{-7}$, about
two times smaller than for the $r^{-3}$ model, and $10$ and $20$
times smaller than inferred from the spike events.  The true density profile is
likely bracketed by the Plummer and $r^{-3}$ models.

It should be emphasized that the procedure of setting $f_p=1$ in the
derivation of these maximum optical depths is not entirely
self-consistent.  The fractions $f_{\rm bound}$ and particularly
$f_{\rm free}$, were derived under the assumption that the dynamics
are dominated by the stellar component.  If this is not the case, then
both the encounter time $\tenc$ and relaxation time $\trelax$ may be
considerably larger.  Precise predictions of the behavior of the
system in the presence of a substantial planetary component by mass
would require a more sophisticated treatment than that presented here.
However, based on the determination of the mass-to-light ratio of M22,
it is certainly true that the core of M22 cannot be currently be
dominated by planetary-mass bodies.  Therefore, the conclusions of
\S\S 4.1-4.3 are secure: essentially regardless of its past dynamical
evolution, a substantial population of planets cannot exist in
the core of M22.  As shown in Figure 3, within a projected radius of
$1'$ from the cluster center, the halo contributes an optical depth of
$\tau_{\rm halo}= 3 \times 10^{-6}$ for the fiducial density model.
This is also true for the Plummer model.  Thus, reasonable models can
reproduce the observed optical depth only if the entire halo is
composed of planets.  This is essentially excluded by measurements of
the surface brightness (Trager et~al.\ 1995) and mass-to-light ratio
(Peterson \& Cudworth 1994) of M22.

\section{Galactic Free-Floating Planets?}

The results of \S 4 strongly suggest that the spike events cannot be
due to microlensing by lenses associated with M22.  Therefore, if one
were to continue with the interpretation that these events are due to
microlensing, and assuming that the direction of M22 constitutes a
generic line of sight toward the Galactic bulge, the lenses constitute
a free-floating Galactic population.  What do the observations of SM22
imply about such a population under this assumption?

The optical depth inferred is independent of the location and nature
of the lenses, therefore $\tau_{\rm spike}=(2.9\pm 1.2) \times
10^{-6}$, as before.  The upper limit on the mass of the lenses is now
less certain, but assuming the lenses have typical bulge distances and
kinematics, I find $M_p \la 0.5\mjup$ -- still firmly in the planetary
regime.  The lower limit on the separation from the lack of detection
of the primary is $a\ga 6.3~\au$.  Thus the planets must still have
relatively wide orbits to escape detection if they are bound to parent
stars.  The MACHO collaboration has measured the optical depth toward
the Galactic bulge based on two different analysis methods.  They find
$\tau=(2.0 \pm 0.4) \times 10^{-6}$ centered at
$(l,b)=(3\deg.9,-3\deg.8)$ using clump giants as sources (Popowski
et~al.\ 2001), and $\tau= (3.2 \pm 0.5) \times 10^{-6}$ centered at
$(2\deg.68,-3\deg.35)$ using difference image analysis (Alcock
et~al.\ 2000).  Note that these optical depths are {\it mutually
exclusive} from that implied by the spike events, as the standard
analysis techniques are not sensitive to events with timescales $\la
1~{\rm day}$.
The optical depth is a relatively strong
function of position. The models of Han \& Gould (1995) indicate that
the optical depth at the position of M22 should be $3-4$ times smaller
than the values measured at the positions reported by MACHO.  
These  estimates are necessarily 
model dependent, however,  for definiteness, I will assume that the optical
depth toward the MACHO fields and Baade's window, $(l=1^\circ,b=-4^\circ)$,
is three times smaller than toward M22.  Thus the six
spike events imply an additional optical depth of
$\tau_{\rm BW}=3\tau_{\rm spike}=9\times 10^{-6}$.  
Assuming that these planetary lenses have similar
spatial and kinematic distributions as the bulge microlenses, this
corresponds to $> 1800 (M_p/0.5\mjup)^{-1}$ planets per bulge
microlens.

These planets cannot be part of a halo population.   The expected contribution
to the optical depth toward M22 from
a standard, spherical, singular, isothermal halo is (e.g., Sackett \& Gould 1993) 
\be
\tau_{\rm halo}=5\times 10^{-7} \left ({v_\infty \over 220~\kms}\right)^2
\label{eqn:tauhalo}
\ee
where $v_\infty$ is the asymptotic halo circular speed.  Introducing a 
core to the halo would decrease this estimate.  Even if only one of the spike
events was due to microlensing by a halo object, this would imply 
an optical depth in planetary-mass objects toward the Large Magellenic Cloud in conflict with
combined EROS and MACHO limits (Alcock et~al.\ 1998). 

These planets also cannot be part of the thin or thick disk.  The optical
depth to an exponential disk is (Gould, Miralda-Escud{\' e} \& Bahcall 1994)
\be
\tau_{\rm disk}={2\pi G \Sigma_d \dos^2 \over h c^2} g(\alpha),
\label{eqn:taudisk}
\ee
where $\Sigma_d$ is the local surface density of the disk, 
\begin{eqnarray}
&\alpha=\dos(|b|/h - 1/R_d),\\
&g(\alpha)=\alpha^{-2}[1-2\alpha^{-1}+(1+2\alpha^{-1})\exp(-\alpha)],
\label{eqn:gfuncs}
\end{eqnarray}
and 
$h$ and $R_d$ are the scale height and scale length of the disk,
respectively.  I adopt $R_d=2.5~\kpc$ (see Sackett 1997, and
references therein).  The optical depth is maximized for this value of
$R_d$ and the line of sight to M22 ($b=-7\deg.6$) for $h=0.6~{\rm
pc}$.  The total observed surface density of matter is $40~\msun~{\rm
pc^{-2}}$ (Gould, Bahcall, \& Flynn 1996, Zheng et~al.\ 2001), while
the surface density of all matter between $\pm 1.1~\kpc$ and the plane
is $71~\msun~{\rm pc}^{-2}$ (Kuijken \& Gilmore 1991), leaving room
for an additional $31~\msun~{\rm pc}^{-2}$ of dark matter between $\pm 1.1~\kpc$ and
the plane.  This corresponds to $\Sigma_d=37~\msun~{\rm pc}^{-2}$, and
thus the maximum optical depth toward M22 from a disklike component is
$\tau_{\rm disk}=4.3\times 10^{-7}$, about a factor of 7 smaller than
that implied by the spike events.

Finally, Binney, Bissantz, \& Gerhard (2000) argue that the optical
depth to resolved events measured by MACHO toward the Galactic bulge
using difference image photometry, $\tau=(3.2 \pm 0.5) \times 10^{-6}$,
is {\it already} difficult to reconcile with our knowledge of Galactic
structure (but see Sevenster \& Kalnajs 2001).  
Thus there is no room for the additional
contribution of $\tau_{\rm BW}\sim 9 \times 10^{-6}$, which would be required
if the free-floating planets were smoothly distributed.

These arguments are, of course, indirect.  However, there are several
ways of unambiguously detecting or ruling out such a free-floating
planet population.  The (unpublished) results of the MACHO spike
analysis toward the Galactic bulge would likely answer this question
definitively.  Performing spike analyses on the OGLE-I or OGLE-II
databases (Udalski et~al.\ 1993; Udalski, Kubiak, \& Szymanski 1997;
Udalski et~al.\ 2000) might prove more difficult, due to the lack of
contemporaneous color information.  However, during the OGLE-III phase
the sampling rates will be increased for some fields, enabling the
resolution of considerably shorter time scale events.  Finally, some
events should be present in the databases of the follow-up
microlensing collaborations.  For example, the PLANET collaboration
(Albrow et~al.\ 1998) monitored $N_e\sim 60$ events toward the
Galactic bulge with median sampling intervals of $\sim 1~{\rm hour}$
(Albrow et~al.\ 2001, Gaudi et~al.\ 2002), sufficient to resolve essentially all events
caused by planets capable of producing the M22 spike events.  The
number of expected events in the PLANET database due to the implied
free-floating planet population can roughly be estimated as
\be
N_{\rm exp} =  {2 N_e \ave{N_*} \tau_{\rm BW} \ave{\Delta t} \over\pi  \ave{\te}},
\label{eqn:nexp}
\ee
where $\ave{N_*}\sim 10^3$ is the average number additional stars on
each frame, $\ave{\Delta t}\sim 40~{\rm days}$ is the average duration
that each event was monitored, and $\ave{\te}\la 1~{\rm day}$ is the
average planetary event timescale.  Thus $N_{\rm exp}\ga 15$ events should
be present in the PLANET database if the implied population of
Galactic free-floating planets is real.

\centerline{{\vbox{\epsfxsize=10.5cm\epsfbox{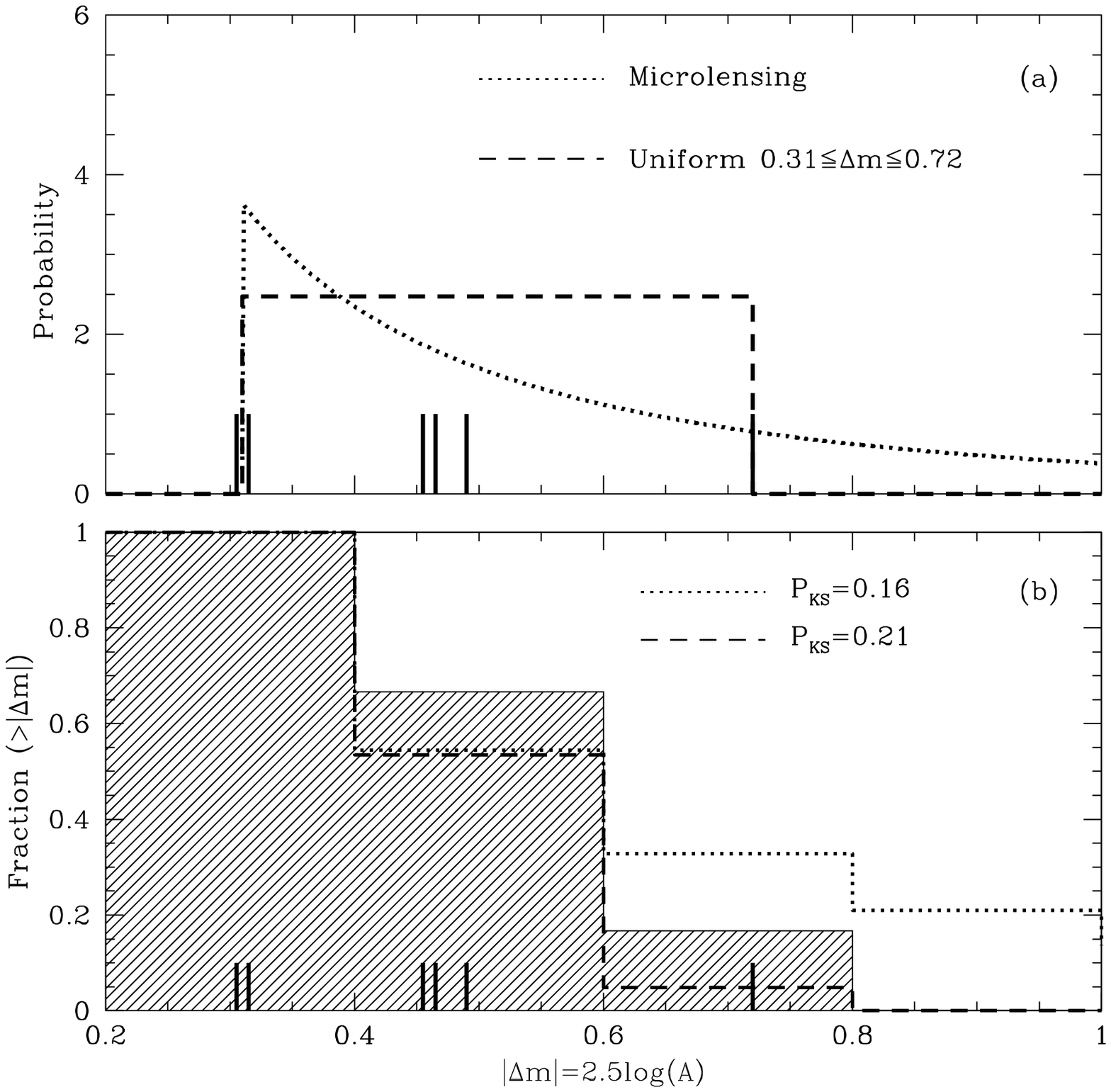}}}}
{\footnotesize {\bf FIG. 4}
(a) Differential distributions of the magnitude difference
$|\Delta m|$ for microlensing (dotted line) and a distribution that is
uniform over the observed range of $|\Delta m|$ (dashed curve).  The curves
are arbitrarily normalized.  The vertical segments are the observed $|\Delta m|$ for the six spike events.   (b) The cumulative distributions 
of $|\Delta m|$, normalized to $|\Delta m|=0.32$ (or $A=1.34$).  The 
shaded histogram is for the six observed events, the dotted histogram is the expected microlensing cumulative distribution, and the dashed histogram is 
for the uniform distribution.  The KS-statistic probabilities are indicated.
}
\bigskip

All of these constraints can be avoided if the planets are not
smoothly distributed.  As suggested by de la Fuente Marcos \& de la
Fuente Marcos (2001), a dark cluster of planets along the line of
sight toward M22 (but not associated with M22) could reproduce the
observed spike events without violating any of the limits on Galactic
structure.  The minimum mass of the dark cluster required to reproduce the observed
optical depth is, 
\be 
M_{\rm dc,min}\simeq 7 \times 10^4 \msun {x \over 1-x}, 
\label{eqn:dc}
\ee 
assuming the cluster radius subtends an angle $\ge
2'\hskip-2pt.5$.  Thus, unless the cluster is very close to us ($\dol \la
1\kpc$), the dark cluster must be quite massive, $M_{\rm dc} \ga
10^4\msun$.  See de la Fuente Marcos \& de la Fuente Marcos (2001) for
a more thorough discussion of the dark cluster interpretation.

\section{Discussion}

The arguments of the previous sections lead to the conclusion that the
only explanation for the origin of the spike events that is consistent
with all available observations and theory is a dark, massive
structure composed of light $M\la 1\mjup$ compact objects
coincidentally along our line of sight to M22.  Such an explanation
seems ad hoc at best.  The simplest alternative is that these events
are not due to microlensing.  While it is not my intention to
attempt to provide a definitive explanation for
the spike events, in light of the evidence against the microlensing
interpretation, it seems worthwhile to review and reassess the reasons
why SM22 favored the microlensing interpretation over other sources of
variability.

The arguments in favor of the microlensing interpretation are
that (1) the light curves are
constant except for the spike event, (2) the colors and magnitudes of
the sources are consistent with an unbiased sample of stars in the CMD
and (3) the distribution of magnifications is consistent with that
expected from microlensing.  The first reason is compelling, and
rejects a large fraction of known variables.  However it is important
to note that the SM22 observations are restricted to
25 epochs over the course of 100 days, and thus only exclude
variables with duty cycles $\ga 4\%$.  The fact that the distribution of source star
magnitudes and colors is consistent with an unbiased sample of stars
may not be very constraining, because the number of proposed
microlensing sources is quite small.  Thus subtle biases are
difficult detect.  Also, it is possible, perhaps likely, that low-mass
main sequence stars are those that are most likely to exhibit the
required behavior: light curves that are constant to a few percent
$>96\%$ of the time with brief brightenings of $\sim 50\%$ for $<4\%$
of the time.  If this were the case one would naturally expect the
distribution of such sources to roughly trace the CMD.  Finally the third
argument, that the distribution of magnifications follows that
expected from microlensing, is also not very constraining 
because of small number statistics.  Figure 4 shows the cumulative distribution
of $|\Delta m|$ for the six spike events. 
For microlensing, the cumulative distribution of magnifications 
normalized to $A=1.34$ is simply $P(>A_T)=\ut^2$, 
where $\ut$ is the radius of the $A=A_T$ magnification contour, and is
given in \eq{eqn:uthresh} for point sources.  The expected
differential and cumulative distributions of $|\Delta m|$ for
microlensing are shown in Figure 4.  Also shown is a distribution
which is uniform in $|\Delta m|$ over the observed range.  Both
distributions describe the data equally well.

At the distance to the Galactic bulge, and $E(B-V)=0.33$, these
sources are {\it brighter} than the theoretical main sequence (Girardi
et~al.\ 2000) by about a magnitude.  They would lie on the main
sequence if at $\sim 5~\kpc$.  Alternatively the bulge stars could be
significantly more reddened than assumed, $E(B-V)=0.73$ versus
$E(B-V)=0.33$.  Note that this would require a substantial amount of
dust along our line of sight between M22 and the bulge.  Without the
original CMD, it is not possible to distinguish between these
scenarios. However given that SM22 conclude that the number
distribution of these source follows the distribution of bulge
sources, the latter scenario seems more likely.  If these sources are
in fact above the bulge main sequence, it is possible that they
sources are blends, e.g.\ M-dwarf/white dwarf binaries.  Such a
scenario may be able to account for the color and magnitude of these
sources, and furthermore may explain the variability as well.

\section{Conclusions}

The primary conclusion of this study is that there is considerable
circumstantial evidence that the six unresolved 
(``spike'') events detected by SM22 are not due to microlensing,
and therefore SM22 have not detected a substantial population of free-floating planets.    
The chain of logic is as follows:
\begin{description}
\item[{(1)}] Planetary-mass lenses bound to parent stars must be
separated by $\ga 8~\au$ to explain the observations, else the influence of the parent lens would
have been detected.
\item[{(2)}]In the core of M22, the encounter timescale is
$\tenc\approx 4~(a/{\rm AU})^{-2}~{\rm Gyr} $.  Thus all planets with
separations $\ga 0.6\au$ have been ionized by random stellar
encounters over the lifetime of the cluster, $T_0=12~{\rm Gyr}$.
\item[{(3)}]The relaxation timescale in the core of M22 is $\trelax =
0.33~{\rm Gyr}$.  Thus all free-floating planets have evaporated from
the core.
\item[{(4)}]For reasonable assumptions,
the maximum possible optical depth to surviving planets in
the halo of M22 falls considerably short of the observed optical depth
of $\tau \sim 3\times 10^{-6}$.
\item[{(5)}] If smoothly distributed, the mass in free-floating Galactic planets required
to produce the observed optical depth is extremely difficult to
reconcile with current knowledge of Galactic structure.
\end{description}

The only logical alternative is a dark cluster of planets with total
mass $M\ga 10^4 \msun$ that happens to be along our line of sight to M22.
This explanation seems ad hoc at best.

Although there are no obvious alternative astrophysical candidates for
these spike events,
considering the weight of the arguments presented here, it would seem prudent
to study existing data in order to
fully characterize stellar variability at the relevant levels.
Similar databases with higher
temporal sampling, such as the HST time series
photometry of the globular cluster 47~Tuc (Gilliland et~al.\ 2000), 
may be able to address this question definitively, without requiring
additional resources.

\acknowledgements

I would like to thank Anthony Aguirre, Arlin Crotts, Xiaohui Fan, and
Pawan Kumar for helpful discussions, John Bahcall for comments and
encouragement, and Eugene Chiang, Andrew Gould, Ivan King, and an
anonymous referee for comments and suggestions that led to an improved
manuscript.  This work was supported by NASA through a Hubble
Fellowship grant from the Space Telescope Science Institute, which is
operated by the Association of Universities for Research in Astronomy,
Inc., under NASA contract NAS5-26555.

{}


\end{document}